\documentstyle[pre,twocolumn,aps]{revtex}
\begin{document}

\draft

\title{Braid analysis of (low-dimensional) chaos}

\author{Nicholas B. Tufillaro\cite{nbt}}
\address{Departments of Mathematics and Physics\\
Otago University, Dunedin, New Zealand}

\date{10 July 1993}
\maketitle
\begin{abstract}
We show how to calculate orbit implications (based on Nielsen-Thurston
theory) for horseshoe type maps arising in
chaotic time series data.
This analysis is applied to data from the Belousov-Zhabotinskii reaction and
allows us to
(i) predict the existence of orbits of
arbitrarily high periods from a finite amount of time series data,
(ii) calculate
a lower bound to the topological entropy,  and
(iii) establish a ``topological model'' of a system  directly
from an experimental time series.
\end{abstract}


\narrowtext
\section{Introduction}
Braids arise as periodic orbits in dynamical systems modeled by
three-dimensional flows  \cite{art1,boy1,mat1,mel1}.
The existence of a single periodic orbit of a dynamical
system can imply the coexistence of many other periodic
orbits. The most well-known example of this
phenomenon occurs in the field of
one-dimensional dynamics and is described by
Sarkovskii's Theorem \cite{sa}.
Less well-known is the fact that analogous results
hold for two-dimensional systems \cite{mat1}.
In one-dimensional dynamics it is useful to study
the period (or the permutation) of an orbit \cite{bald}.
In  two-dimensional systems it is useful to
study the {\it braid type} of an orbit \cite{boy1}.
Given this specification, we can ask whether or not the
existence of a given braid (periodic orbit) {\it forces} the existence of
another; as in the one-dimensional case, algorithms have
recently been developed for answering this question \cite{bes,los,franks}.

As originally observed by Auerbach et.\ al., unstable periodic orbits are
available in abundance from a single chaotic time series using the
method of close recurrence \cite{aur,nbtrel}.
By a ``braid analysis'' we propose to analyze a chaotic
time series by first extracting an (incomplete) spectrum
of periodic orbits, and second
ordering the extracted orbits according to their
orbit forcing relationship.
As shown in this paper,
it is often possible to find a single periodic orbit (or a small
collection of orbits) which forces many orbits in the observed
spectrum. This single orbit also forces additional
orbits of arbitrarily high period.
This analysis is restricted to ``low-dimensional'' flows (roughly, flows
which can be modeled by systems with one unstable Lyapunov exponent),
however it has a strong predictive capability.

We would also like to point out that this analysis gives us an
effective and mathematically well defined ``pruning procedure'' for
chaotic two-dimensional diffeomorphisms \cite{pred1}. Instead of asking for
rules describing which orbits are missing (pruned), we instead look for
those orbits which must be present. All orbits of a low enough period (say  up
to
period 10)  appear to be succesfully perdicted by
this method. This procedure will usually miss orbits of
higher period, however from an experimental viewpoint the low period
orbits are the most important and accessible.
Orbits of low period often force an infinity of other orbits.
This is illustrated in
one-dimensional dynamics by the famous statement ``period three implies
chaos'' \cite{li}. An analogous statement in two-dimensional dynamics is that a
non-well-ordered period three braid implies chaos \cite{gam1}.

This paper is organized as follows. In section II we briefly review the theory
of orbit forcing in one- and two-dimensional maps (three-dimensional flows).
In section III we show how the
braid analysis method works by  applying it to times series data generated from
the Rossler equations. This section also discusses some useful
numerical and symbolic refinements to the method of close recurrence.
In section IV we apply the braid analysis method to data from
the Belousov-Zhabotinskii reaction. Our analysis builds directly on
the original topological analysis of this data set due to
Mindlin et.\ al.\ \cite{mind1}. In section V we offer some concluding remarks
by
indicating how this method can be extended to low-dissipation and conservative
systems.
In the examples studied in this paper
we
do have good control of the symbolics.
In principle, though, this method does not require good control of the
symbolics (a good partition) and can thus overcome some of the current
difficulties associated with finding good symbolic descriptions
for (nonhyperbolic) strange attractors \cite{gras}.

\section{Orbit Forcing Theory}

In this paper we study in detail the orbit structure of
horseshoe type attractors. In particular, we  will make a good deal of
use of those features of one-dimensional dynamics
which must always carry over to two dimensions.
In our review we closely follow the point of view
adopted in the Ph.D thesis of T.\ D.\ Hall \cite{toby}.
Thus, none of the results in this theoretical section are new.
However
they are presented in such a way as to make them more useful
for those wishing to apply these theoretical results to experimental
data sets, as is done in the later sections of this paper.
Let us begin by summarizing some well-established \cite{bald,nit}
definitions
and results pertaining to the periodic orbit structure of maps of the line.
Let
$F:{\bf R} \rightarrow {\bf R}$
be a continuous map,
and
$R = \{x_1,\dots,x_n\}$
be a period $n$ orbit of $F$ with
its points labeled in such a way that
$x_1 < x_2 < \cdots < x_n$.
Then there is a cyclic permutation
$\pi = \pi(R,F)$
associated to $R$, defined by
$F(x_i) = {x_{\pi(i)}}$.
A relation
$\leq_1$
on the set $C$ of all cyclic permutations can
be constructed as follows:
if
$\sigma, \tau \in C$,
then $\tau \leq_1 \sigma$
if and only if
every map
$F:{\bf R} \rightarrow {\bf R}$
which
has a periodic orbit $R$ with
$\pi(R,F) = \sigma$
also has a
periodic orbit $S$ with
$\pi(S, F) = \tau$.
If this is the case then we shall say that the permutation
$\sigma$
{\it forces}
the permutation
$\tau$.
The relation
$\leq_1$ is a partial order which can be calculated,
in the sense that there is a simple algorithm to determine whether
or not
$\tau \leq_1 \sigma$
for any two cyclic permutations
$\tau$ and $\sigma$:
unfortunately this algorithm takes a long time
to execute for orbits of high period, and it is difficult to use it
to analyze the global structure of the partially
ordered set
$(C, \leq_1)$ \cite{bald}.
There is, however, an interesting subset
of $C$ onto which the restriction of $\leq_1$ is well-understood.
Let
$\text{UM} \subseteq C$
be the set of
{\it unimodal permutations}: that is,
the set of permutations of periodic orbits of the {\it tent map}
(or one-dimensional horseshoe) defined well into the
parameter regime where a complete horeshoe exists
(a full shift on two symbols) without prunning
\begin{displaymath}
           F(x) =
            \left\{
             \begin{array}{ll}
                3x & 0 \leq x \leq 1/2\\
                3(1-x)& 1/2 \leq x \leq 1\\
                0&{\text{otherwise}}
             \end{array}
           \right.
\end{displaymath}
(equivalently, UM is the set of permutations which can be realized by
periodic orbits of unimodal maps whose turning point is a maximum).
The partial order
$\leq_1$
restricts to
a linear order on UM which is described by kneading theory \cite{mil}.
Indeed, this observation leads to a statement of topological
universality in the one-dimensional
context: there is a universal order in which the periodic
orbits are built up in families of unimodal maps of the line, and
this order is calculable using kneading theory.

A periodic orbit $R$ of a homeomorphism of the plane can be described
by its {\it braid type} \cite{boy1}; the reader who lacks the mathematical
background
necessary to appreciate the following definitions should simply regard
the braid type as a two-dimensional analog
of the permutation, and the term `pseudo-Anosov' as one which, while
essential for the accurate statement of theorem 1, is not essential for
applying
the results found
in later statements. Very roughly, braids can be divided into three types, the
so-called {\it finite order braids} --- braids which only force the
existence of a finite number of other braids (periodic orbits), ---
{\it pseudo-Anosov braids} --- braids which must force the existence
of an infinite number of other braids, and
{\it reducible braids} --- braids which can be decomposed into
distinct components which fall into one of the first two types.
Pseudo-Anosov braids play an important
role in the study of chaotic three-dimensional flows
since, at the topological level, they force the
coexistence of an infinite number of other periodic orbits.
The term pseudo-Anosov arises within the context of the Nielsen-Thurston
classification of isotopy classes of surface automorphisms, which is the
main tool used in the proofs of the results presented: the reader
interested in this powerful theory could consult Ref.\ \cite{thur}, in which
there is
also an extensive bibliography. An exposition of the relevance
of Nielsen-Thurston theory to two-dimensional dynamics, with
particular reference to braid types, can be found in Ref.\ \cite{boy1}.

Let $f$ and $g$ be orientation-preserving homeomorphisms of the plane which
have periodic orbits $R$ and $S$ respectively.
We say that $(R, f)$ and $(S, g)$ {\it have the same braid type}
if there is an orientation-preserving homeomorphism
$h:{\bf R}^2\setminus R \rightarrow {\bf R}^2\setminus S$
such that
$h \circ f|_{{\bf R}^2\setminus R} \circ h^{-1}$
is isotopic to
$g|_{{\bf R}^2\setminus S}$.
Having the same braid type is an equivalence relation on the set of
pairs $(R,f)$, and we write bt$(R,f)$ for the
equivalence class containing $(R,f)$, the
{\it braid type} of the periodic orbit $R$ of $f$.
We say that the braid type bt$(R,f)$ is of pseudo-Anosov,
reducible, or finite order type, according to the Nielsen-Thurston
type of the isotopy class of
$f : ( S^2, R  \bigcup \{\infty \}) \rightarrow
(S^2, R \bigcup \{ \infty \})$
(see Ref.\ \cite{hand} for a statement of Thurston's
classification theorem for homeomorphisms relative to a finite
invariant set). By analogy with  the one-dimensional theory, we define
a relation
$\leq_2$
on the set BT of all braid types as follows:
if
$\beta, \gamma \in \text{BT},$
then
$\gamma \leq_2 \beta$
if and only
if every homeomorphism of the plane which has a periodic orbit
of braid type
$\beta$
also has a periodic orbit of braid type
$\gamma$.
If this is the case, we shall say that the braid type $\beta$
{\it forces} the braid type $\gamma$.
The relation $\leq_2$ is a partial order \cite{boy2}:
as in the one-dimensional case, there is an algorithm for calculating
this order (the so-called ``train track algorithm'') \cite{bes},
but it is difficult to use it to analyze the global
structure of the partially ordered set (BT, $\leq_2$).
Still, using a version of this
algorithm due to
Bestvina and  Handel \cite{bes},
it is possible to
calculate the orbits forced by an individual braid up to
say period 10 or so by hand, and we have done this for
several low period pseudo-Anosov horseshoes braids in order
to find their associated topological entropies (see Table \ref{table1}).
Also, an automated version of this algorithm promises to extend these
calculations to higher period orbits \cite{tad}.

We restrict our attention to the subset HS $\subseteq$ BT of {\it horseshoe
braid types}: that is, those which are realized by periodic orbits of the
horseshoe map $f$ \cite{smale}.
The determination of constraints on the order in which periodic orbits can
be built up in the creation of horseshoes is tantamount to the analysis
of the partially ordered set (HS, $\leq_2$); and an important observation in
discussing the topological universality of two-dimensional systems
is
that, unlike (UM, $\leq_1$), the partially ordered set
(HS, $\leq_2$) {\it is not linearly ordered}.

The connection between the ordered sets
(UM, $\leq_1$)
and
(HS, $\leq_2$)
is that periodic orbits of both the tent map and the
horseshoe have a symbolic description.
We associate to points in the non-wandering set of each map
an infinite sequence of $0's$ and $1's$ in the
usual manner, and describe a period
$n$
orbit
$R$
by its
{\it code}
$c_R$,
given by the first
$n$ symbols of the sequence associated to the rightmost point of $R$.
We shall identify periodic orbits of the two maps which have
the same code, using the same symbol $R$ to denote each orbit.
Such an identified pair $R$ will be referred to as a
{\it horseshoe orbit} (or simply as an {\it orbit}):
the fact that the orbit is periodic will always be assumed.
We shall write
$c^\infty_R$ for the semi-infinite sequence
$c_R c_R c_R \ldots$ (which is associated to the
rightmost point of $R$, regarded as a periodic orbit of the tent map);
and
$^\infty c^\infty_R$
for the bi-infinite sequence
$\ldots c_R c_R c_R \ldots$
(which is associated to the rightmost point $R$, regarded  as a periodic
orbit of the horseshoe).

It will be convenient for us to make another identification.
If $R$ is a period $n$ orbit, then denote by
$\tilde c_R$ the code obtained by changing the last symbol
of $c_R$.
If $\tilde c_R$ is not the two-fold repetition of a sequence of
length $n/2$, then it is the code of a period $n$ orbit
$R'$ which can be shown to have the same permutation and braid
type as $R$.
We identify $R$ and $R'$ in what follows, using the same symbol
$R$ to denote both, referring to them as a single orbit.
The code $c_R$ of the identified pair will always be chosen to end with
a zero (this is to ensure that the sequence $c^\infty_R$ contains the group of
symbols
$\ldots  010 \ldots$,
which is necessary for the algorithms which follow).
After making this identification,
distinct horseshoe orbits have distinct permutations:
however, it is possible for several different orbits
to have the same braid type.
With this in mind, we say that
a function defined on the set of
all horseshoe orbits is a
{\it braid type invariant} if it takes
the same value on any two orbits of the same braid type.

Given two horseshoe orbits $R$ and $S$, we shall
write $S \leq_1 R$ if and only if
$\pi(S,F) \leq_1 \pi(R,F)$; and we shall write
$S \leq_2 R$ if and only if
$\text{bt}(S,f) \leq_2  \text{bt}(R,f)$.
We refer to $\leq_1$ and $\leq_2$ as the one- and
two-dimensional {\it forcing orders} respectively.
Our aim is to analyze $\leq_2$ using the well-understood properties of
$\leq_1$.
Notice that the orders $\leq_1$ and $\leq_2$ cannot strictly
be regarded as being defined on the same set:
if $R$ and $S$ are two horseshoe orbits which have the same
braid type but different permutations, then they must be regarded
as being equal when considering the order $\leq_2$, but as being
distinct when considering the orbit $\leq_1$. For example,
the orbits
$f_{3 \times 2}$
and $s^2_6$ in Table \ref{table1} are
two such orbits
(for an example analogous example in
the orientation-reversing Henon map see Ref.\ \cite{hansen}).

We say that a horseshoe orbit $R$ is
{\it quasi-one-dimensional} (qod) if for all
orbits $S$ we have $R \geq_1 S \Longrightarrow R \geq_2 S$.
As proved in
Ref.\ \cite{toby}, the following result provides a simple
classification of the qod orbits
which have pseudo-Anosov braid type.
Let $\hat {\bf{Q}}$ denote the set
${\bf{ Q }} \cap (0,1/2)$ of rationals lying (strictly)
between 0 and 1/2. Given such a rational
$q=m/n$ (in lowest terms), we write
$P_q$ for the period $n+2$ orbit which has code
$c_q = 10^{\kappa_1(q)}1^20^{\kappa_2(q)}1^2 \ldots 1^2 0^{\kappa_m(q)} 1 0$,
where $\kappa_1(q) = \lfloor1/q \rfloor -1$,
and
$\kappa_i(q) = \lfloor i/q \rfloor - \lfloor (i-1)/q \rfloor -2$
for $2 \leq i \leq m$
(here $\lfloor x \rfloor$ denotes the greatest integer which does not exceed
$x$).
Thus, for example, $c_{2/5} = 1011010$. Then we have the following
result of T.\ D.\ Hall \cite{toby}:
\\
\\
\noindent
{\bf Theorem 1} $q \longleftrightarrow P_q$  is a one-to-one correspondence
between $\hat {\bf Q}$ and the set of quasi-one-dimensional horseshoe orbits
which have pseudo-Anosov braid type. Moreover
$P_q \geq_2 P_{q'} \Longleftrightarrow q' \geq q$ for all
$q,q' \in \hat {\bf Q}$.
\\
\\
\noindent
It is a consequence of quasi-one-dimensionality that
$P_q \geq_2 P_{q'} \Longleftrightarrow P_q \geq_1 P_{q'}$, and
therefore the braid types of the orbits $P_q$ form a linearly
ordered subset of BT. The final statement of the theorem tells us
that the ordering within this subset is simply the
reverse of the usual ordering on $\hat {\bf Q}$.

Theorem 1 allows us to quickly identify a very useful subset of pseudo-Anosov
braids
and a  program is easily written to quickly
generate
all the horseshoe qod orbits of any desired period.
Moreover, Theorem 1 can be used to define an invariant of
horseshoe braid type: the {\it height}
$q(R)$ of a horseshoe orbit $R$ is the unique element of
$[0,1/2]$
with the property that
for all
$q \in \hat {\bf Q}$
we have
$q < q(R) \Longrightarrow P_q \geq_1 R$
and
$q > q(R) \Longrightarrow P_q \not \geq_1 R$
(thus $q(R) = \text{sup}\{ q\ \in \hat {\bf Q} : P_q \geq_1 R\}$).
Because the orbits $P_q$ are qod, it follows
immediately that
$q<q(R) \Longrightarrow P_q \geq_2 R$. In fact, it can be shown
that the second property in the definition of
height also holds for the two-dimensional forcing order: that is \cite{toby},
\\
\\
\noindent
{\bf Theorem 2}
Let $R$ be a horseshoe orbit and
$q \in \hat {\bf Q}$.
Then $q < q(R) \Longrightarrow P_q \geq_2 R$ and
$q > q(R) \Longrightarrow P_q \not \geq_2 R$.
In particular, height is a braid type invariant.
\\
\\
We can also
find the finite order braids
because the finite order braids
of the horseshoe
are exactly those whose period equals the denominator of the height
\cite{toby,hol}.
Because the height is defined in terms of the one-dimensional order,
it can be calculated using kneading theory.
We first define the height $q(c) \in (0,1/2]$ of a semi-infinite
sequence which begins $c = 10\ldots,$ and which contains the group
of symbols $\ldots 010 \ldots$.
To do this, write $c$ in the form
$c = 10^{\kappa_1} 1^{\mu_1} 0^{\kappa_2} 1^{\mu_2} \ldots$,
where $\kappa_i \geq 0$ and
$\mu_i$ is either 1 or 2 for each $i$,
with $\mu_i = 1$ only if $\kappa_{i+1} > 0$ (thus the $\kappa_i$ and
$\mu_i$ are uniquely determined by $c$, and the fact that $c$
contains the group $\ldots 010 \ldots$ means that
$\mu_s = 1$ for some  $s$). Let
\begin{displaymath}
I_r(c) = \left
(  {r \over { {2r+\sum_{i=1}^r } \kappa_i} },
{r \over {(2r-1) + \sum_{i=1}^r \kappa_i} }
\right  ]
\end{displaymath}
for each $r\geq1$, and let $s$ be the least positive integer such that either
$\mu_s = 1$ or $\cap_{i=1}^{s+1} I_i(c) = 0$.
Then $\cap_{i=1}^s I_i(c) = (x,y]$ for some
$x$ and $y$: we define
$q(c) = y$ if $\mu_s = 1$ or $I_{s+1} (c) > \cap_{i=1}^s(c)$,
and
$q(c) = x$  if $\mu_s = 2$ and $I_{s+1}(c) < \cap_{i=1}^s I_i(c)$.
\\
\\
\noindent
{\bf Theorem 3} $q(R) = q(c_R^\infty)$: in particular, $q(R)$ is
positive and rational \cite{toby}.
\\
\\
For example,
let $R$ be the period 20 orbit with code
$c_R = 10000110001100001100$.
We have
$\kappa_1 = 4$, $\mu_1 = 2$,
$\kappa_2 = 3$, $\mu_2 = 2$,
$\kappa_3 = 4$, $\mu_3 = 2$,
$\kappa_4 = 2$, and
$\mu_4 = 1$.
Thus
$I_1 = (1/6,1/5]$,
$I_2 = (2/11, 2/10]$,
$I_3 = (3/17, 3/16]$, and
$I_4 = (4/21, 4/20]$.
Since $4/21 > 3/16$
we have
$I_1 \cap I_2 \cap I_3 \cap I_4 = 0$,
and
$I_4 > I_1 \cap I_2 \cap I_3$.
Therefore
$q(R) = \text{max}(I_1 \cap I_2 \cap I_3 ) = 3/16$.

There is a relationship between height and a well-established braid
type invariant: the height of a horseshoe is equal to the left hand
endpoint of its rotation interval.
A  proof of this is given in Ref.\ \cite{toby}, where a practical algorithm
for determining the rotation interval of a horseshoe is described.
It can also be shown using the algorithm for determining the height
that for each $q = m/n$ (in lowest terms),
$P_q$ is the only period $n+2$ orbit of height $q$;
thus it is the only orbit of its braid type. This
observation enables us to determine exactly which of the orbits
$P_q$ are forced by a given orbit R in the two-dimensional order:
we now present an algorithm for this purpose.
If $R$ is a horseshoe  orbit then we define the {\it depth}
$r(R) \in (0,1/2] \cap {\bf Q}$ of $R$ as follows:
consider all groups of the form $\ldots 01110 \ldots$ or
$\ldots 01010 \ldots$ in the sequence
$^\infty c_R^\infty$; suppose
that there are $l$ such groups
$g_1, \ldots , g_l$ contained in one period of $^\infty c^\infty_R$.
If $l = 0$ then
$r(R) = 1/2$. Otherwise, for each
$i \leq l$ let $f_i$ be the code obtained by starting at the last
$1$ in $g_i$ and moving forwards through $^\infty c_R^\infty$, and $b_i$ be
that obtained by starting at the first $1$ and moving backwards. Then
$r(R) = \text{min}_{1\leq i\leq  l} \text{max} (q(f_i), q(b_i))$.
\\
\\
\noindent
{\bf Theorem 4}
$r(R)$ is the unique element of $[0,1/2]$ with the property that for
all $q \in \hat {\bf Q}$ we have $r(R) < q \Longrightarrow R \geq_2 P_q$
and
$r(R) > q \Longrightarrow R {\not \geq_2} P_q$. In particular, depth is a
braid type invariant \cite{toby}.
\\
\\
The following corollaries from theorems 2 and 4  will be very useful
in our  braid analysis since they allow to us to apply one-dimensional
theory to calculate  two-dimensional orbit forcings.
\\
\\
{\bf Corollary 1}
Let $R$ and $S$ be horseshoe orbits.
If $r(R) < q(S)$ then $R \geq_2 S$. On the
other  hand,
if $q(R) < q(S)$ and
$r(S) < r(R)$ then
$R {\not \geq_2} S$  and $S {\not \geq_2} R$: thus orbits of
the braids types $R$ and $S$ can exist independently of each other.
\\
\\
In fact, a much stronger result can be proved:
if $r(R) < q(S)$ then every homeomorphism of the plane
which has a periodic orbit of braid type $R$ has at least as many
periodic orbits of the braid type $S$ as does the horseshoe (the
corollary only says that every such homeomorphism has at least
one such orbit). Thus all of the periodic orbits of the braid type
$S$ must be created before any of the periodic orbits of braid
type $R$ in any family of homeomorphisms leading to the creation
of a horseshoe.
We can use the following corollary to locate some of these orbits.
\\
\\
{\bf Corollary 2 }
If a homeomorphism of the disc has a qod orbit
$P_q$, and $R$ is another orbit which is
forced by $P_q$ in the one-dimensional
order (which means that its height is $\geq q$), then the
homeomorphism must have
at least one orbit of the braid type of $R$. If in addition the height
of $R$ is {\it strictly} greater than $q$, then the
homeomorphism must have at least
as many orbits of the braid type of $R$ as does the horseshoe \cite{toby2}.
\\
\\
For example, if a period 7 orbit with the braid type of
$1011010$ (height 2/5) is extracted, then there must be
at least one orbit of the
braid type of $10110$
(since this braid type has height 2/5), and at least
as many of the braid type of 101110 as in the horseshoe (i.e., 2)
need exist (since the height of these orbits is also $2/5$). On
the other hand, both of the orbits in the period 6 pair $10111^0_1$
must
exist since the height of these orbits are $1/2$.
A listing of the low period qod orbits and some of the
low period orbits which they force is found in Table \ref{table3}.

Since the qod orbits essentially inherit the one-dimensional
forcing order we can use one-dimensional methods to calculate their
topological entropies.
This information can be used to obtain lower bounds for
the topological entropy of a partially-formed horseshoe.
More precisely, given a horseshoe orbit $R$,
let $h(R)$ denote the smallest possible
entropy of a homeomorphism having an orbit of the same
braid type as $R$.
Now we know that for all $q > r(R)$ we have $R \geq_2 P_q$,
and hence the topological entropy $h(R) \geq h(P_q)$.
However, since $P_q$ is a qod orbit, it can be shown that
$h(P_q)$ is equal to the entropy of $P_q$ regarded as
an orbit in one dimension: this can be calculated using standard
transition matrix techniques.
A {\it Mathematica} program based on the method of
Block et.\ al.\ \cite{block} is available from
T.\ D.\ Hall
which allows us to calculate the topological
entropies of the qod orbits.
A graph of the function $(q, h(P_q))$ appears in Ref.\ \cite{toby};
it is monotonic and discontinuous everywhere. Thus, we can
go directly from a qod orbit to a lower bound for the
topological entropy. For example, the orbit $R$
with code $c_R = 10011010$ has $r(R) = 1/3$: any
partially formed horseshoe which includes an orbit of the braid type of $R$ has
entropy
greater than $h(P_{2/5}) \approx 0.442$.
By taking
rationals closer and closer to $1/3$ we get better and better estimates
(e.g., 5/14 gives 0.481).
We can also use the pseudo-Anosov orbits which are not qod to get
estimates (e.g., the orbit $s_8^4$ mentioned above has $h \approx 0.498$),
however in this
instance
we must compute the entropy from a train track and this
can be a  difficult calculation.

In Table \ref{table1} we collect together some useful facts about
braids in the horseshoe up to period eight.
All the reducible orbits up to period nine
only have finite order components.
This is because the lowest period with a  pseudo-Anosov
orbit is five, and numbers less than or equal to nine
do not have proper factors greater than or equal to five.
As mentioned previously, the topological entropy
for the pseudo-Anosov orbits  which are not quasi-one-dimensional
is calculated by finding the ``train track'' using the
method of Bestvina and Handel \cite{bes}. This method results not
only in a topological entropy,
but also an explicit Markov partition
(a `topological model')
and
associated Perron-Frobenius matrix which is useful for
locating in phase space
where the predicted
periodic orbits are to be found.

\section{Rossler Braid Analysis}
A braid analysis of a low dimensional chaotic
time series consists of four steps once
an appropriate three-dimensional space is created \cite{mind1}:
(i) the periodic orbits are extracted by the
method of close recurrence \cite{book,sch}, (ii) the braid type
of each periodic orbit is identified and the orbits are
ordered by their two-dimensional forcing relationship \cite{bes,toby}
(iii) a subset of braids are selected which have maximal
forcing and which force the orbits extracted in step (i),
and (iv)
if possible, an attempt is made to
verify that some of the predicted
orbits (not originally extracted in step (i)) are found in
the system.

In practice, steps (i) and (ii) are greatly simplified if
the {\it template} or
{\it knot-holder} organizing the flow can be identified using the
procedure described by Mindlin and co-workers \cite{mel1,mind1,mind2,nbtnmr}.
Knowledge of the template helps in obtaining the
symbolic names of the periodic orbits and in calculating
the forcing relationship for the specific braids in that template.
For instance, if the template is identified as a two-branch
horseshoe knot holder (as are all the examples
studied in this paper), then the theory of qod orbits
of section II can be applied to simplify the analysis.

Although template identification is very valuable, it
is not essential for a braid analysis. Nor is the symbolic
identification of the extracted orbits. In the worst case
a braid
analysis does require that the the braid conjugacy class of
each extracted periodic orbit is identified
(see Elrifai and Morton \cite{mor},
or Jaquemard \cite{jaq} for algorithms), and that the minimal Markov model
(a `train track' in the language of
Thurston)
can be constructed for
each braid (see Bestvina and Handel \cite{bes}, Los \cite{los}, and Franks and
Misiurewicz \cite{franks} for algorithms). Algorithms exist for both
of these steps, although the most computationally efficient
version of the braid conjugacy algorithm is probably not
an effective solution beyond $B_8$.

To illustrate the braid analysis we consider
a chaotic attractor of the Rossler equations,
\begin{eqnarray*}
\dot x &=& -(y+z)\\
\dot y &=& x + ey\\
\dot z &=& f + xz -\mu z
\end{eqnarray*}
with $e = 0.17$, $f=0.4$ and $\mu = 0.85$.
The Rossler equations are integrated through $10^5$ cycles and the
return map is examined at the half plane
$\Sigma = \{ (x,y,z): x < 0\ \& \  y = 0\}$. The
template is easily identified as a horseshoe with zero global torsion.
This template identification is verified by calculating the relative
rotation rates and linking numbers of the extracted periodic orbits
as described by Mindlin et.\ al.\ \cite{mind1,book,mind2}.

To  extract the (surrogate) periodic orbits by the method
of close recurrence we first convert the  return map from
the sequence of values $(x_n, z_n)$ directly into
the symbol sequence of $0's$ and $1's$. In this particular
instance, since the map is close to one-dimensional,
a good symbolic partition is obtained by examining the maximum
value of the next return map formed from the projection on
the $x$-coordinate --- $(-x_n, -x_{n+1})$ --- at the surface of section.
Orbits passing to the left of the maximum are
labeled zero, and those to the right
are labeled one. Next we search this symbolic encoding for each
and every periodic symbol string. Every time a periodic symbol string
is found we calculate its recurrence and then save the instance of the
 orbit with
the best
recurrence. For instance, in searching for the period
three orbit `100' we search the symbolic encoding of the
return map for any instance of `100' and its cyclic permutations
`010' and `001', and every time this symbol string is found,
we next calculate its recurrence, which for this period three
orbit is $\epsilon_{y=0} = (x_{n+3}-x_{n})^2 + (z_{n+3}-z_{n})^2$, and
then save the orbit with the minimum $\epsilon$. The advantage
of this procedure of orbit extraction is that it is exhaustive.
We search for every possible orbit up to a given period. In
these studies we searched for all orbits between periods
1 and 16.

The resulting spectrum of periodic orbits up to period eight is
shown in Table \ref{table4}.  The orbits which are present in
(the full shift) complete hyperbolic system, and not present in Table
\ref{table4}, are
said to be {\it pruned}. Our goal is to predict as best as
possible the pruned spectrum from the chaotic time series.

Before we discuss the braid analysis, though, it is interesting to
consider the number of orbits extracted as a function
of the number of points in the return map.  As expected, we find that
the number of orbits
that can be extracted
increases with the number of points in the return map.
More importantly, our numerical results strongly suggest that
using the method of close recurrence
it is possible to obtain all the low
period periodic orbits embedded within the
strange attractor after examining only a finite number of data points,
in this instance $10^5$.
For instance, after $10^4$ points are examined our numerical results show that
no
new periodic orbits are found below period six. Similarly,
after examining $10^5$ points we believe we have found all
orbits up to period eight.
These results also caution us when working with
small data sets ---  the extracted orbit spectrum is expected
to miss orbits either because the orbit is pruned (it is not
in the strange set) or because the sample of the strange set
we are examining
fails in providing a close enough coverage
over the whole attractor.

Using the results in section II, the finite-order and
quasi-one-dimensional orbits in the extracted spectrum are
easily identified from their orbit codes (see Table \ref{table4}).
Not unexpectedly,
we
see a sequence of quasi-one-dimensional orbits of
increasing entropy (decreasing height) --- the
maximal qod orbit  is the period 16 orbit 1011011011011010 with
entropy $h \approx 0.480804$.
All orbits forced by this period 16 orbit up to period 8 are
present, and none are missing.
So this period 16 qod orbit already gives us a very good
hyperbolic set with which to approximate to our (possibly nonhyperbolic)
chaotic attractor. Can we do better? Doing better in this
instance means identifying a pseudo-Anosov orbit which is
not qod, but perhaps implies the maximal qod orbit found.
Indeed, in this data set there is such an orbit, it is
the period 8 orbit 10010100 with entropy $h \approx 0.498093$.
Again, the spectrum of orbits forced by this maximal
pA orbit are consistent with the extracted spectrum which was
examined up to
period 16.

This data set is close to one-dimensional so
kneading theory also does quite well for predicting
the low period orbits. For instance, we could consider the
period 3 orbit
100, and this
orbit (based on one-dimensional unimodal theory) also accurately predicts
most of the extracted spectrum.
However, this period 3 orbit is finite order, and we know
from the results of Holmes and Whitley that there will be
many (possibly high period) orbits which are forced by $\leq_1$ but
not by $\leq_2$ \cite{whit}
(in fact, we know that in 2-dimensions 100 forces only itself and a fixed
point, and
in 1-dimension it forces orbits of arbitrarily high period).
Thus, although one-dimensional theory is a useful
guide in this instance to the low period orbits,
it can not be safely applied to
make  predictions about high period orbits.

\section{Belousov-Zhabotinskii Reaction Braid Analysis}

To further illustrate the braid analysis method we
obtained data from the Belousov-Zhabotinskii chemical reaction \cite{bz}.
This is the same data set analyzed by
Mindlin et.\ al.\ and consists of 65,000 equally spaced points
which measure the time dependence of the bromide ion concentration
in the stirred chemical reactor.
Following the techniques described by  Mindlin et.\ al.\ we
also embedded the scalar time series $x(i),\ i = 1, 2, \ldots, N$
in ${\bf R}^3$ via a {\it differential phase space}  embedding
described by
\begin{eqnarray*}
y_1(i) &=& x(i) + \lambda * y_1(i-1), \ \ \lambda = 0.995\\
y_2(i) &=& x(i)\\
y_3(i) &=& x(i) - x(i-1)
\end{eqnarray*}
from which we reproduced the 3-dimensional attractor and return map
shown in Figs.\ 5 and  6(a) of Ref.\ \cite{mind1}.
The attractor is a zero global torsion horseshoe.
There are approximately 125 points per cycle so the return
map consists of about 520 points.

Our technique for extracting (surrogate) periodic orbits from
this time series differs somewhat from that described in Ref.\ \cite{mind1}.
We use the same procedure described for the Rossler data:
first the return map data is converted into a symbol sequence of $0's$ and
$1's$ depending on whether the orbit passes to the left or right
of the maximum value of the return map,
and second an exhaustive search is performed
for all possible periodic orbits between periods 1 and 15.
Again, this data set
is almost one-dimensional and the simple  symbolic prescription just described
leads to a unique and consistent encoding of all the periodic orbits
we are able to extract.
Since the data set (at the return map) is small, we choose not
to pick an arbitrary cut off for $\epsilon$, the close
return criterion. Rather, we report the best $\epsilon$ we are able to extract
for a given periodic orbit (see Table \ref{table6}).
By including this additional piece of information we can make
a more selective judgement about which close returns
are, and are not, good surrogates for periodic orbits. In this
way we are able to locate a few more periodic orbits than were
originally reported in Ref.\ \cite{mind1}.
Also, one perhaps surprising result comes out of this extraction method.
It is not uncommon to find orbits of high period with very small
recurrences. For example the period 13 orbit 1011011101110 has
a recurrence of $\epsilon = 0.000712$ which is significantly better
than almost all of the lower period orbits.

A list of the extracted orbits and their Thurston types is
presented in Table \ref{table6}.
As expected, there is a sequence of quasi-one-dimensional orbits
of increasing entropy, the largest of which is the
period 16 orbit 1011011011011011 with entropy $h \approx 0.48084$.
In this particular instance, the period 16 qod orbit is in fact
the maximal pseudo-Anosov orbit in the data set and it forces all the extracted
orbits in this data set  except for the finite order period
three orbit $101$. Indeed, a careful analysis of
this data set suggests that
the signal is subject to a small parametric drift which carries
it between the strange attractor and a stable period 3 orbit
(whose 1-dimensional entropy is $h \approx 0.4812$) \cite{gil}.

Table \ref{table7} shows the number of predicted and extracted orbits
as a function of period. The
number of forced orbits which are not in the extracted data
set increases with the period. As with the Rossler data set,
we believe the forced orbits which are missing
could actually be extracted from
the data set if we were given a longer time series which provides
a better coverage of the entire attractor.

\section{Conclusion}

In retrospect, we find it remarkable that such a small
subset of periodic orbits (which are rather
easy to get from experiments) contain so much topological and
dynamical information about a (low-dimensional) flow.
As previously demonstrated,
a few low period orbits are sufficient
to determine the template describing the stretching and folding
of the strange set \cite{mel1,mind1,mind2}. The template provides an upper
bound
to the topological entropy and is, in a sense, a maximally (i.e., a full shift)
hyperbolic set which we formally associate to a (possibly
nonhyperbolic) strange set \cite{book}. In this paper we show how a sequence
of periodic orbits (and their associated hyperbolic sets) can be used
to obtain a collection of finer and finer  approximations
to a strange set which is probably not hyperbolic. Formally, we
might say that the hyperbolic set associated with each pseudo-Anosov braid
is embedded within the strange attractor we are trying to describe in the
sense that the (possibly nonhyperbolic) strange set
must contain at least all the orbits forced by the
extracted pseudo-Anosov braid. We indirectly discussed two
measures of the goodness of this approximation --- the difference in
the topological entropy, and the difference between the forced
and extracted low period orbits. Using either measure, we have seen
that it is possible to select moderate period  orbits
(say $<$ period 20) which provide good hyperbolic sets
with which we can approximate
a (possibly nonhyperbolic) strange set.

The  dynamical information derived from an orbit depends only on its
braid type. As mentioned in section III, the braid type of an orbit
can be determined without  obtaining a good symbolic description of the
orbits in the flow. We will illustrate these techniques in
a future paper in which we consider a braid analysis of the bouncing
ball system in a low-dissipation regime \cite{bbb}.
Our strategy for handling the cases
where a good symbolic description is not easy
to obtain is to find a complete set of braid invariants on the small
set of braids of interest. For instance, as mentioned
in Table \ref{table1}, the exponent sum (simply the sum of the crossings)
is a complete braid type invariant for horseshoe braids up to period 8.

We would also like to point out that it is common for a
collection of finite order braids to be pseudo-Anosov.
This suggests an alternative strategy: instead finding a
single orbit with maximal implications, it should also
be possible to find a collection of orbits
(possibly with very low period)
which force all the observed orbits and provides yet another
hyperbolic approximating set. This is also the subject
of future investigations.

\acknowledgements
We wish to thank Bob Gilmore whose questions began this
investigation,
Toby Hall for carefully explaining his
results prior to publication, Tad White
for providing us with his train track program which
we used, again with Toby's help and Paul Melvin's, to
verify the entropy results presented in
Table 1, and Eric Kostelich for providing the
Belousov-Zhabotinskii data file.

\onecolumn
\mediumtext

\begin{table}
\caption{Low period horseshoe orbits.
The exponent sum is a complete braid type invariant for all
horseshoe orbits up to period eight (see Ref. [17] for
the explicit conjugacy).
The braid name notation is from Ref.\ [35].
Thurston braid types: finite order (fo), reducible (red), and pseudo-Anosov
(pA),
and quasi-one-dimensional (qod).
The ``red, fo'' orbits are reducible Thurston types with finite order
components.
\label{table1}
}
\begin{tabular}{|c|l|c|c|c|c|c|}
$P$ & $c_P$ & $\pi_P$ & $Type$ & $es$ & $q(P)$ & $h$  \\
braid name& symbolic name& permutation & Thurston type&
exponent sum& height & topological entropy \\ \tableline
$s_1$ & $1$ & (1) & fo & 0 & 1/2 & 0                 \\ \hline
$s_2$ & $10$ & (12) & fo & 1 & 1/2 & 0               \\ \hline
$s_3$ & $10_1^0$ & (123) & fo & 2& 1/3 & 0          \\ \hline
$f_{2 \times 2}$ & $1011$ & (1324)& red, fo & 5 &1/2& 0   \\
$s_4$ & $100^0_1$ & (1234) & fo & 3& 1/4 & 0         \\ \hline
$s^1_5$ & $1011^0_1$ & (13425)&fo&8&2/5&0     \\
$s^2_5$ & $1001^0_1$& (12435)&pA, qod&6&1/3&0.543535\\
$s^3_5$&$1000^0_1$&(12345)&fo&4&1/5&0              \\ \hline
$s^1_6$&$10111^0_1$&(143526)&red, fo&13& 1/2&0        \\
$f_{3\times2}$&100101&(135246)&red, fo&9&1/3&0           \\
$s^2_6$&$10011^0_1$&(124536)&red, fo&9&1/3&0        \\
$s^3_6$&$10001^0_1$&(123546)&pA, qod&7& 1/4&0.632974\\
$s^4_6$&$10000^0_1$&(123456)&fo&5&1/6&0            \\ \hline
$s^1_7$&$101111^0_1$&(1453627)&fo&18&3/7 &0         \\
$s^2_7$&$101101^0_1$&(1462537)&pA, qod&16&2/5&0.442138\\
$s^3_7$&$100101^0_1$&(1362547)&pA&14&1/3&0.476818  \\
$s^4_7$&$100111^0_1$&(1254637)&pA&14&1/3&0.476818  \\
$s^5_7$&$100110^0_1$&(1356247)&fo&12& 2/7&0         \\
$s^6_7$&$100010^0_1$&(1246357)&pA&10& 1/4&0.382245   \\
$s^7_7$&$100011^0_1$&(1235647)&pA&10&1/4&0.382245   \\
$s^8_7$&$100001^0_1$&(1234657)&pA, qod&8& 1/5&0.666213\\
$s^9_7$&$100000^0_1$&(1234567)&fo&6&1/7&0          \\ \hline
$f_{{2^2} \times 2}$&$10111010$&(15472638)&red, fo&23& 1/2 & 0\\
$s^1_8$&$1011111^0_1$&(15463728)&red, fo&25&1/2&0  \\
$s^2_8$&$1011011^0_1$&(14725638)&fo&21&3/8&0       \\
$s^3_8$&$1001011^0_1$&(13725648)&pA&19&1/3&0.346034\\
$s^4_8$&$1001010^0_1$&(13647258)&pA&17&1/3&0.498093\\
$s^5_8$&$1001110^0_1$&(13657248)&pA&17&1/3&0.498093\\
$s^6_8$&$1001111^0_1$&(12564738)&pA&19& 1/3&0.346034\\
$s^7_8$ & $1001101^0_1$ & (12573648) & pA & 17 &1/3& 0.498093\\
$f_{2 \times 4}$ & $10001001$ & (13572468) & red, fo&13&1/4&0\\
$s^8_8$&$1000101^0_1$&(12473658)&pA&15& 1/4& 0.568666 \\
$s^9_8$& $1000111^0_1$ & (12365748) & pA &15& 1/4 & 0.568666 \\
$s^{10}_8$&$1000110^0_1$&(12467358)&red, fo&13& 1/4 &0\\
$s^{11}_8$&$1000010^0_1$&(12357468)&pA&11&1/5&0.458911\\
$s^{12}_8$&$1000011^0_1$&(12346758)&pA&11&1/5&0.458911\\
$s^{13}_8$&$1000001^0_1$&(12345768)&pA, qod&9&1/6&0.680477\\
$s^{14}_8$&$1000000^0_1$&(12345678)&fo&7&1/8&0\\
\end{tabular}
\end{table}

\narrowtext

\begin{table}
\caption{A sequence of
{\it quasi-one-dimensional} (qod)
orbits with increasing forcing implications (entropy).
Thus,
every orbited forced by the period 7 qod orbit is also forced by
the period 10 orbit, and so on. The notation ``(*)'' after an orbit
indicates that one (but perhaps not both) of the saddle-node pair
is forced. Both saddle-node partners
will be forced by the
next orbit in the qod sequence.\label{table3}}
\begin{tabular}{|c|l|l|l|}
      & Forced by (7)& Forced by (10)&Forced by (13)\\
Period&$101101_1^0$&$101101101^0_1$&$101101101101^0_1$\\ \tableline
1 & & & \\ \hline
2&10&&\\ \hline
3&&&\\ \hline
4&1011&&\\ \hline
5&$1011^0_1$(*)&&\\ \hline
6&$10111^0_1$&&\\ \hline
7&$101111^0_1$&$101101^0_1$&\\ \hline
8&$10111010$&$1011011^0_1$(*)&\\
 &$1011111^0_1$&&\\ \hline
9&$10111111^0_1$&$10110111^0_1$& \\
 &$10111101^0_1$&&\\
 &$10110101^0_1$(*)&&\\ \hline
10&$101111111^0_1$&$1011010111$&$101101101^0_1$\\
 &$101111101^0_1$&$101101101^0_1$&\\
 &$101110101^0_1$&&\\
 &$101101111^0_1$(*)&&\\ \hline
11&$1011111111^0_1$&$1011010111^0_1$&$1011011011^0_1$(*)\\
  &$1011111101^0_1$&$1011010101^0_1$&\\
  &$1011110111^0_1$&&\\
  &$1011110101^0_1$&&\\
  &$1011011111^0_1$(*)&&\\
  &$1011011101^0_1$(*)&&\\ \hline
$\vdots$&$\vdots$&$\vdots$&$\vdots$\\
\end{tabular}
\end{table}

\begin{table}
\caption{Periodic orbits extracted from time series
data of the Rossler equations. All extracted orbits up to
period 8 are shown. Between periods 9 and 16 only the extracted
quasi-one-dimensional ({\it qod}) are shown.
The maximal pseudo-Anosov orbit ($1001010^0_1$) is not qod,
but it forces all qod orbits with height greater than $1/3$.
Thurston Types: finite order (fo),
reducible (red), and pseudo-Anosov (pA).\label{table4}}
\begin{tabular}{|l|l|c|l|l|}
Period& Name& Height& Entropy& Type  \\ \tableline
1&       1&1/2& 0 &fo  \\ \hline
2&       10&1/3& 0 &fo  \\ \hline
3&       101&1/3& 0 &fo  \\
3&       100&1/3& 0 &fo  \\ \hline
4&       1011&1/2& 0 &red  \\ \hline
5&       10111&2/5&0 &fo  \\
5&       10110&2/5&0 &fo  \\ \hline
6&       101110&1/2&0 &red  \\
6&       101111&1/2&0 &red  \\
6&       100101&1/3&0 &red  \\ \hline
7&       1011111&3/7&0 &fo  \\
7&       1011110&3/7&0 &fo  \\
7&       1011010&2/5&0.442138 &pA, qod\\
7&       1011011&2/5&0.442138 &pA, qod\\
7&       1001011&1/3&0.476818 &pA  \\
7&       1001010&1/3&0.476818 &pA  \\ \hline
8&       10111010&1/2&0 &red  \\
8&       10111110&1/2&0 &red  \\
8&       10111111&1/2&0 &red  \\
8&       10110111&3/8&0 &fo  \\
8&       10110110&3/8&0 &fo  \\
8&       10010110&1/3&0.346034 &pA  \\
8&       10010111&1/3&0.346034 &pA  \\
8&       10010101&1/3&0.498093 &pA\\
8&       10010100&1/3&0.498093 &pA\\ \hline
9&       101111010&3/7&0.397081 &pA, qod\\
9&       101111011&3/7 &0.397081&pA, qod\\ \hline
10&      1011011010&3/8&0.473404&pA, qod\\
10&      1011011011&3/8&0.473404&pA, qod\\ \hline
11&      10111111010&4/9&0.373716&pA,  qod\\
13&      1011011011010&4/11&0.479450&pA, qod\\
13&      1011011011011&4/11&0.479450&pA, qod\\ \hline
15&      101111111111011&6/13&0.354176&pA, qod\\
15&      101101111011010&5/13&0.467734&pA, qod\\ \hline
16&      1011011011011010&5/14&0.480804&pA, qod\\
\end{tabular}
\end{table}

\begin{table}
\caption{Periodic orbits extracted from the
Belousov-Zhabotinskii Reaction time-series
up to period 15.
All orbits with a best (normalized) recurrence of less than 0.1 are shown.
The period 16 orbits are from Ref.\ [15]. \label{table6}}
\begin{tabular}{|l|l|l|l|}
Period& Name& Recurrence& Type  \\ \tableline
1&       1&               0.016782& fo\\ \hline
2&       10&              0.002615&fo\\ \hline
3&       101&             0.000128&fo\\ \hline
4&       1011&            0.002648&fo\\ \hline
5&       10111&           0.002962&fo\\
5&       10110&           0.013668&fo\\ \hline
6&       101110&          0.006449&fo\\
6&       101111&          0.029014&fo\\ \hline
7&       1011110&         0.005088&fo\\
7&       1011010&         0.041837&pA, qod, $h = 0.442138$\\
7&       1011011&         0.010585&pA, qod, $h = 0.442138$\\ \hline
8&       10111010&        0.014287&fo\\
8&       10110111&        0.017370&fo\\
8&       10110110&        0.070293&fo\\ \hline
9&       101111010&       0.001720&pA, qod, $h = 0.397081$\\
9&       101101011&       0.061843&pA\\
9&       101101010&       0.020884&pA\\
9&       101101110&       0.018380&pA\\
9&       101101111&       0.050094&pA\\ \hline
10&      1011101010&      0.037910&fo\\
10&      1011101011&      0.007237&fo\\
10&      1011111010&      0.035966&fo\\
10&      1011011010&      0.003440&pA, qod, $h = 0.473404$\\
10&      1011011011&      0.008971&pA, qod, $h = 0.473404$\\ \hline
11&      10110101010&     0.038044&\\
11&      10110111010&     0.009119&\\
11&      10110110111&     0.008447&\\
11&      10110110110&     0.085593&fo \\ \hline
12&      101110101011&    0.031644&\\
12&      101110101010&    0.013333&\\
12&      101101010110&    0.029674&\\
12&      101101111010&    0.055765&\\
12&      101101101011&    0.032222&\\
12&      101101101010&    0.016560&\\
12&      101101101110&    0.048563&\\
12&      101101101111&    0.064420&\\ \hline
13&      1011110101010&   0.052853&\\
13&      1011010111010&   0.022397&pA\\
13&      1011010101110&   0.004291&\\
13&      1011011101110&   0.000712&\\
13&      1011011101010&   0.011199&pA\\
13&      1011011111010&   0.046255&pA\\
13&      1011011010110&   0.097165&pA, qod, $h=0.479450$ \\
13&      1011011011010&   0.013063&pA, qod, $h=0.479450$ \\
13&      1011011011011&   0.004663&\\ \hline
14&      10110101110110&  0.056515&\\
14&      10110101010110&  0.009366&\\
14&      10110111010110&  0.047037&\\
14&      10110110111010&  0.060024&\\
14&      10110110110110&  0.004920&fo\\ \hline
15&      101101011101110& 0.034250&\\
15&      101101110101010& 0.015669&\\
15&      101101101011010& 0.010945&\\
15&      101101101010110& 0.076061&\\
15&      101101101110110& 0.049261&pA\\
15&      101101101111010& 0.007018&\\
15&      101101101101010& 0.049573&\\
15&      101101101101110& 0.010284&\\
15&      101101101101111& 0.078043&\\ \hline
16&      1011011011101010&na&pA\\
16&      1011011011111010&na&pA\\
16&      1011011011011010&na&pA, qod, $h=0.480804$\\
16&      1011011011011011&na&pA, qod, $h=0.480804$\\
\end{tabular}
\end{table}

\begin{table}
\caption{Number of periodic orbits extracted
and predicted for
time series data from
the Belousov-Zhabotinskii reaction ($\approx$ 500 points in
the return map). As illustrated with the
Rossler data, we expect that the time series is
far to short to be able to extract all the predicted orbits of any
except the lowest periods.
\label{table7}}

\begin{tabular}{|c|c|c|c|c|}
Period&\# predicted& \# found&\# missing&\# marginal\\
\tableline
1&1&1&&\\ \hline
2&1&1&&\\ \hline
3&1&1&&\\ \hline
4&1&1&&\\ \hline
5&2&2&&\\ \hline
6&2&2&&\\ \hline
7&4&3&1&\\ \hline
8&5&2&2&1\\ \hline
9&8&4&2&2\\ \hline
10&11&5&6&\\
\end{tabular}
\end{table}

\end{document}